\documentclass[11pt]{article}
\usepackage{amsmath,amsfonts,amsthm,amsbsy}
\usepackage{graphicx}
\def\be{\begin{equation}}
\def\ee{\end{equation}}

\def\bq{\tilde q}
\def \sech{{\rm sech}}

\begin{document}

 \title{The exact evaluation of the corner-to-corner resistance
   of an $M\times N$ resistor network: \\Asymptotic expansion}
 \vskip 1cm
\author{J. W. Essam\\ Department of Mathematics\\
 Royal Holloway College, University of London\\
 Egham, Surrey TW20 0EX, England \\
    \\
F. Y. Wu \\ Department of Physics \\
Northeastern University\\
 Boston, Massachusetts 02115, U.S.A.}

 \date{}
 \maketitle

 

\abstract{We study the  corner-to-corner resistance of an $M\times N$ resistor
network with resistors $r$ and $s$ in the two spatial directions,
and obtain an asymptotic expansion of its exact expression
for large $M$ and $N$.  For $M=N$, $r=s=1$, our result is
\begin{align}
R_{N\times N} = \frac 4 \pi \log N + 0.077\,318 +\frac {0.266\,070}{N^2}
 -\frac {0.534\,779} {N^4} + O(\frac 1 {N^6}) .\nonumber
\end{align}   }
 
\section{Introduction}
A classic problem in the theory of electric circuits is the computation of
the resistance between two nodes in a resistor network. Formulated by
Kirchhoff [\ref{kirch}] more than 160 years ago, the problem has been studied by
numerous authors over many years (see, for example, [\ref{pol}, \ref{ds}]). Kirchhoff
  explored the graph-theoretical aspect of the algebraic formulation
and obtained  the two-point resistance
in terms of 2-rooted spanning forests and
spanning trees.
But the formulation, while elegant, does not provide sufficient physical insights.
Past studies have instead focused
 on infinite networks for which analysis can be carried to fruition [\ref{cserti}].

The computation of the asymptotic expansion of the
corner-to-corner resistance of a rectangular resistor network
has been of interest for some time,
as its value  provides
 a lower bound to the resistance of
 compact percolation clusters in the Domany-Kinzel model
of a directed percolation [\ref{DK}]. The  corner-to-corner
resistance has been studied   by one of us (JWE) numerically using
the method of a differential approximants [\ref{GU}] together with a
Neville table analysis [\ref{ETB}].

 Recently, one of us (FYW) has re-visited the two-point resistance problem [\ref{wu}], and
deduced a closed-form expression for the resistance between arbitrary two nodes
for finite networks.  However,  the exact expression obtained in [\ref{wu}] is  in the
form of a double summation whose mathematical and physical contents
 are not immediately apparent. In this paper, we
take a closer look at this summation formula and obtain its asymptotic expansion
for large lattices.

The organization of this paper is as follows: In Sec. 2 we recall
the expression of the corner-to-corner resistance in an $M\times N$
resistor network obtained in [\ref{wu}], and reduce it to a form
more manageable for our purposes. 
One of the two summations in the resistance expression is carried out in Sec. 3
by using a new summation identity which we derive. The resulting expression is  written
in the form of   a dominant term plus a correction.
Asymptotic expansions of the dominant and correction terms are obtained in Secs. 4 and 5,
and we summarise the results in Sec. 6.  We also show that the exact
expression of the asymptotic expansion  is in agreement
with those determined numerically  [\ref{ETB}].  

\section{Formulation of the summation formula}
 Consider a rectangular $M\times N$ network of resistors with resistances $r$ and $s$
 on edges of the network in the respective horizontal and vertical directions. For definiteness,
we  consider both $M,N$ even, and expect the asymptotic expansion to be independent of
this choice. The example of an $M=6, N=4$ network is  shown in Fig. 1.
 \begin{figure}[htbp]
\begin{center}
\includegraphics[scale=0.4]{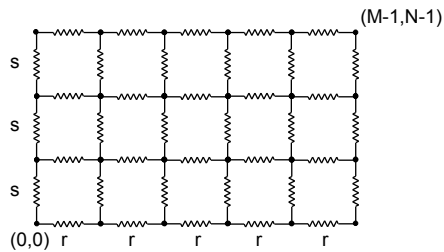}
  \caption{\it An $M\times N$ resistor network.}
   \label{fig:resist}
 \end{center}
 \end{figure}

Using Eq. (37) of
[\ref{wu}], the resistance between opposite corner nodes $(0,0)$ and $(M-1, N-1)$
of the network is
\begin{align}
&R_{\{M\times N\}}(r,s) = \frac {r(M-1)} {N}  + \frac {s(N-1)} {M}   \nonumber\\
&+\frac 2 {MN}{\sum_{m=1}^{M-1}\sum_{n=1}^{N-1}
\frac {\Big[\cos \big( \frac 1 2 {\theta_m} \big) \cos \big( \frac 1 2 {\phi_n} \big)
 - \cos\big(M-\frac 1 2\big)\theta_m \cos\big(N-\frac 1 2\big)\phi_n
\Big]^2 }
{r^{-1} (1-\cos \theta_m )+s^{-1}(1-\cos \phi_n)  } } \label{exact}
 \end{align}
where $\theta_m= {m\pi}/M, \phi_n= {n\pi}/ N.$
Re-arranging the numerator  in the summand,  (\ref{exact}) becomes
\begin{align}
&R_{M\times N}(r,s) = \frac{r(M-1)}N + \frac{s(N-1)}M\nonumber\\
&+\frac 8{MN}{\sum_{m=1}^{M-1}\sum_{n=1}^{N-1}}_{(m+n
\,\hbox{odd})}\frac{\cos^2(\theta_m/2)\cos^2(\phi_n/2)}
{r^{-1}(1-\cos \theta_m)+s^{-1}(1-\cos
 \phi_n)}\label{Rexact}
\end{align}

There are two possibilities for the restriction $m+n =$ odd to hold, namely,
\begin{align}
& m= 2p-1, n=2q, \quad p=1,2,.., M /2, \ q = 1,2,..., N /2, \nonumber \\
& n= 2p-1, m=2q, \quad p=1,2,..,N /2, \,\ q = 1,2,...,  M /2. \nonumber
\end{align}
Splitting the sum into two
parts accordingly and introducing the notation
\begin{align}
A_q = &\frac {q\pi} N, \quad B_p = \Big( p- \frac 1 2\Big)  \frac \pi {M}, \nonumber \\
\intertext{we obtain}
 &R_{M\times N}(r,s) = (rs)^\frac12[R_{M\times N}(r/s) +R_{N\times M}(s/r)] \label{finalexpansion} \\
\intertext{where}
 R_{M\times N}(\rho)& =\frac{{\sqrt \rho} (M-1)}N \nonumber \\
& + \frac{4\sqrt \rho}{MN}\sum_{p=1}^{ M/
 2}\sum_{q=1}^{ N /2} \bigg[ \frac
{\cos^2A_q(1+ \rho\sin^2A_q)}{\rho \sin^2A_q +\sin^2B_p} -\cos^2 A_q \bigg]. \label{rho}
\end{align}

Sums of the term $\cos ^2A_q$ can be carried out  using the identity
\begin{equation}
\sum_{q=1}^{N/2} \cos^2 \Big( \frac {q\pi} N \Big) = \frac N 4 - \frac 1 2.
\end{equation}
This yields
\begin{align}
R_{M\times N}(\rho)&=\sqrt \rho \left( \frac MN -\frac 12\right) +S_{M\times N}(\rho)\nonumber \\
\intertext{and}
  R_{M\times N}(r,s)
&= \sqrt {rs} \bigg[ \sqrt \rho \left( \frac MN -\frac 12\right) + \frac 1 {\sqrt \rho }
  \left( \frac NM -\frac 12\right) \nonumber \\
  & \qquad \qquad + S_{M\times N}(\rho) + S_{N\times M}(1/\rho) \bigg] \label{R}
  \intertext{where }
  S_{M\times N}(\rho)
 &=\frac{4\sqrt \rho}N\sum_{q=1}^{ N/2}(\cos^2 A_q)(1+
\rho\sin^2A_q)\,S_{q,M,N}(\rho)\label{SqMN}
 \intertext{with}
S_{q,M,N}(\rho) &=
\frac 1{M}\sum_{p=1}^{ M/2}\big[{\rho \sin^2A_q +\sin^2B_p}\big]^{-1} \nonumber \\
&= \frac 1{M} \sum_{k=0}^{(  M /2) -1} \Big[{\rho \sin^2 A_q +\sin^2
\Big(\frac{(k+\frac 12)\pi}{M}\Big)}\Big]^{-1} . \label{Sqdef}
\end{align}
 
\section{Evaluation of $S_{q,M,N}(\rho)$}
It is tempting to evaluate the summation \eqref{Sqdef} 
by using the Euler-Maclaurin summation formula. But as shown in the Appendix
 the Euler-Maclaurin summation is inadequate since it does not determine an error term which  
 cannot be ignored.  We  proceed here to evaluate $S_{q,M,N}(\rho)$
by using a summation identity which we state as a lemma:

\bigskip
\noindent
{\it Lemma:}
\begin{align}
\sum _{k=0}^{(M/2)-1} \frac 1{\rho \sin^2A_q+ \sin^2[(k+\frac 1 2)\frac \pi M]}
   =&\, R(y^*) \nonumber \\
\equiv & \, \frac {M \tanh (\pi y^*)} {2 \sqrt \rho \sin A_q \sqrt{1+\rho\sin^2A_q} }, \label{lemma}
\end{align}
where $M=$ even and $y^*=y^*_{q,M,N}(\rho)$ is defined by
\begin{align}
\sinh \frac {\pi y^*} M =& \sqrt \rho \sin A_q .\label{y}
\end{align}

 \begin{figure}[htbp]
\begin{center}
\includegraphics[scale=0.4]{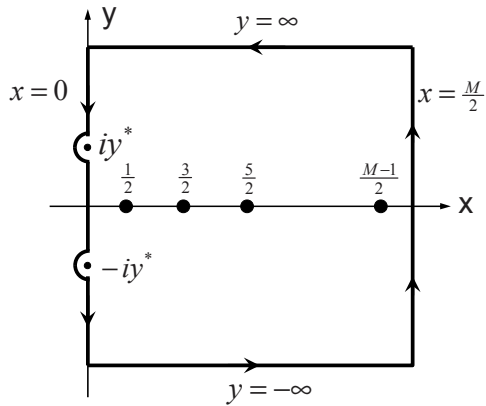}
  \caption{\it Contour of integration $C$ in \eqref{int}. Solid circles
denote simple poles enclosed by $C$.}
   \label{fig:contour}
 \end{center}
 \end{figure}

\noindent
\begin{proof}

Consider the contour integral
\begin{equation}
J_{q,M,N}(\rho) = \frac 1 {2\pi i} \oint_C
\frac {\pi \tan (\pi z) dz}{\sin^2{( \frac {\pi z}  M})+ \rho\sin^2A_q} \label{int}
\end{equation}
where 
the contour $C$ consists of the lines
\begin{equation}
 x= \frac M 2 , \quad
 y= - \infty, \quad y=\infty \label{bound}
\end{equation}
and the imaginary axis $x=0$  with  two half circles of radii $\epsilon\to 0$ around
the two points $z = \pm i y^*$ as shown in Fig. 2.
The contour encloses  $\frac M 2 +2 $ simple poles of the integrand at $z= \pm \,i y^*$ and
$z=\frac 1 2, \frac 3 2, ..., \frac {M-1} 2$.
The residue is $R(y^*)$
 at the simple
poles on the $y$-axis  and \,
$-[{\rho \sin^2A_q+ \sin^2(k+\frac 1 2)\frac \pi M]^{-1}}$\, at\,
$z=k+\frac 1 2,\, k=0, 1, ...$.

  The integration along the contour $C$
vanishes on the lines $y=\pm \infty$, and on the straight line portions of $x=0, \frac M 2$ since
the integrand is odd in $y$. Hence the contour  integral
 is nonzero only on the two half circles. The integrand
is odd in $z$
 so that the integral along the lower
 half circle is equal to the integral in the anti-clockwise direction along
 the reflection of the upper half circle in the $y-$axis. The integral
$J_{q,M,N}(\rho)$ along the contour $C$ may therefore
 be obtained  by integrating round a circle centered on $iy^*$.
 Thus, by the residue
 theorem, the residue at $iy^*$ is equal to the sum of the residues of the $\frac M 2 +2 $ simple poles
enclosed by $C$, hence
   \begin{equation}
R(y^*) = 2 R(y^*) - \sum _{k=0}^{(M/2)-1} \frac 1{\sin^2[(k+\frac 1 2)\frac \pi M]+\rho \sin^2A_q},
\end{equation}
 which yields \eqref{lemma}.
 \end{proof}

\smallskip
The substitution of \eqref{lemma} into \eqref{Sqdef} and \eqref{SqMN}
now yields
\begin{align}
S_{q,M,N}(\rho) &= \frac {\tanh (\pi y^*) } {2 \sqrt \rho \sin A_q \sqrt{1+\rho\sin^2A_q} } \label{SQ} \\
\intertext{and}
  S_{M\times N}(\rho) &=\sum_{q=1}^{ N/2}D_{q,N}(\rho) \tanh (\pi y^*), \label{SMN1}
\end{align}
where 
\begin{equation}
 D _{q,N}(\rho) = \frac 2 N \cdot \frac {\cos^2 A_q \sqrt{1+ \rho \sin^2 A_q} }{\sin
A_q}.\label{Deltadef}
\end{equation}

Anticipating that the dominate  contribution of $ S_{M\times N}(\rho)$ is given by \eqref{SMN1}
with $\tanh (\pi y^*)$ replaced by $1$ (see Appendix), we rewrite \eqref{SMN1} as 
\begin{equation}
 S_{M\times N}(\rho) =S^{(1)} _N(\rho) + \Delta _{M,N}(\rho), \label{SSD}
\end{equation}
where 
\begin{align}
S^{(1)} _N(\rho) &= \sum_{q=1}^{ N/2}D_{q,N}(\rho) \label{s1}
\end{align}
is the dominate contribution, and
\begin{align}
\Delta _{M,N}(\rho)= \sum_{q=1}^{ N/2} \Delta _{q,M,N}(\rho) \label{DeltaMN1}
\end{align}
is the correction with
\begin{equation}
 \Delta _{q,M,N}(\rho) = D_{q,N}(\rho)\big[ \tanh (\pi y^*) -1 \big]. \label{SMN}
\end{equation}

Numerical evaluation of the difference $\Delta_{q,M,N}(1)$ using
$\tanh (\pi y^*)$ given by \eqref{lemma} for $M=N$ and small
values of $q$ shows that it initially decreases with  $N$ but
ultimately shows a rapid increase. For $q=1$ the turning point is
$N=6$ and for $q=2$ it is $N=12$. However $\Delta_{q,M,N}(\rho)$ for fixed
$N$ decreases exponentially with increasing $q$,
 a fact which will
 be seen to hold
for general $M$ and $N$ later (see Eq. \eqref{DeltaResult} below).
The sum in \eqref{DeltaMN1} 
therefore converges rapidly.
 
 The two terms $S^{(1)}_N(\rho)$ and $\Delta_{M,N}(\rho)$
in \eqref{SSD} are evaluated in the next two sections.

\section{Evaluation of $ S_{ N}^{(1)}(\rho)$}
 The asymptotic form of $ S_{ N}^{(1)}(\rho)$ given by the summation (\ref{s1}) is now
  deduced using the Euler-Maclaurin sum
formula ([\ref{Hi}] equation 5.8.13)
\begin{align}
\sum_{p=1}^r f_p&= \frac 1 h \int_{x_0}^{x_r} f(x)dx +\frac 12
\big[f(x_r)-f(x_0)\big] \nonumber\\
&+ \sum_{i=1}^m \frac
{B_{2i}h^{2i-1}}{(2i)!}\Big[ f^{(2i-1)}(x_r)-f^{(2i-1)}(x_0)\Big]+E_m(\eta_m) \label{alternateEM}
\end{align}
where $f_p$ is such that $f_p=f(x_0+p\,h)$, the integer $r$ is finite and the error term is given by
\begin{equation}E_m(\eta_m) = r \frac
{B_{2m+2}h^{2m+2}}{(2m+2)!}f^{(2m+2)}(\eta_m), \qquad
 x_0<\eta_m<x_r. \label{EulerM2}
\end{equation}
But the direct application of ({\ref{alternateEM}) to effect the summation in
 \eqref{s1}  leads to
a divergent integral so we add and subtract $1/A_q$ to the summand
and use (\ref{alternateEM}) with $f(x)$ given by
\begin{equation}\label{fdef}
f(x) \equiv f_\rho(x)= \frac {\cos^2x}{\sin x} \sqrt{1 + \rho
\sin^2x}-\frac 1 x.
\end{equation}
Using $x_0=0, x_r= \pi/2, h = \pi/N, r=N/2$ and since $f_\rho(x)$ does not diverge
at small $x$, the error term $E_m$ is of the order
of $O(N^{-(2m+1})$ and can be  neglected in $m\to\infty$. Denoting by $U_N(\rho)$ and $L_N(\rho)$
the respective correction
to the integral at the upper an lower limits, we obtain
\begin{equation}
  S_{N}^{(1)}(\rho) = I(\rho) + S_N + U_N(\rho) + L_N(\rho)\, , \label{ISUL}
\end{equation}
where $I(\rho)$ is the integral
\begin{align}
I(\rho)&=\frac{2 }\pi \int_0^{\pi/2} f_\rho(x)dx \nonumber \\
  &=\frac{ 1}\pi
\Big[-1 + 4 \log 2  - 2 \log \pi  - \log(1+\rho)
+\frac{\rho-1}{\sqrt \rho} \tan^{-1}\sqrt \rho \,\Big]. \label{I}
\end{align}
The second term in (\ref{ISUL}) is the added summation $S_N$, which
 can be evaluated using the result ([\ref{Hi}] chapter 5, problem 26)
as
\begin{align}
S_N&=\frac{2 }N\sum_{q=1}^{ N/2}\frac 1 {A_q}= \frac{2}\pi\sum_{q=1}^{ N/2}\frac 1 q
 = \frac 2 \pi \Big(\log \frac N 2+\gamma +\frac 1 N -
  \sum_{m=1}^\infty \frac{4^m
B_{2m}}  {2 m N^{2m}} \Big),
 \end{align}
 where $\gamma=0.577\,215\,664\,901\,53\dots$ is Euler's constant.

 The first part of $f_\rho(x)$ is antisymmetric about $\pi /2$ so
the odd derivatives at the upper limit arise entirely from the
$-1/x$ term and is independent of $\rho$.  Hence for $j$ odd $f^{(j)}_\rho(\pi/ 2)
=(-1)^{j+1} j! (2/\pi)^{j+1}$ and the correction to the integral
from the upper limit  is
\begin{equation}\label{U}
U_N(\rho) = \frac 1 N f\Big(\frac \pi 2\Big)+\frac 2 N \sum_{i=1}^m \frac
{B_{2i}h^{2i-1}}{(2i)!}f^{(2i-1)}_\rho\Big(\frac \pi 2 \Big) = \frac{- 2}
{\pi N}+\frac 2 \pi\sum_{i=1}^m \frac {4^iB_{2i}}{2i N^{2i}}
\end{equation}
which, as $m\rightarrow \infty$, cancels terms of the inverse powers of $N$
in $S_N$.

At the lower limit we have $f_\rho(0)=0$ and
\begin{equation}\label{L}
L_N(\rho) = -\frac 2 \pi \sum_{i=1}^m
\frac {B_{2i}}{(2i)!}\left(\frac \pi N
\right)^{2i}f^{(2i-1)}_\rho(0).
\end{equation}
Using  Bernoulli numbers  $B_2= 1/6, B_4 =-
1 /{30}, B_6 = 1/{42}$ ([\ref{Hi}] equation 5.8.8),
the leading terms in $L_N$ are
\begin{align}
L_N(\rho)=&  \frac 2 \pi \Big[ -\frac  {\pi^2} {12
N^2}f^{(1)}_\rho(0) +\frac  {\pi^4} {720 N^4} f^{(3) }_\rho(0)
 -\frac  {\pi^6} {30240 N^6} f^{(5)}_\rho(0) + O\big(\frac 1 {N^8}\big) \Big]
\end{align}
with
\begin{align}\label{fdevs}
f^{(1)}_\rho(0) &=\frac 16(-5 + 3\rho), \nonumber \\
f^{(3)}_\rho(0) &= \frac 1{60}(67 - 210\rho - 45\rho^2), \nonumber \\
f^{(5)}_\rho(0)&=\frac 1{126}(-95 + 3843\rho + 2835\rho^2 +
945\rho^3).
\end{align}
Combining  (\ref{ISUL}) - (\ref{U}), we obtain 
the result
\begin{align}
S^{(1)}_{ N}(\rho) = I(\rho) +\frac 2 \pi
\Big[\log \frac N 2 &+ \gamma \Big]  +L_N(\rho).
 \label{Sum1}
 \end{align}

\section{Evaluation of $\Delta_{q, M,N}(\rho)$}
We now evaluate $\Delta _{M,N}(\rho)$ given by the summation
\eqref{DeltaMN1}   with $\Delta_{q,M,N}(\rho)$ given by \eqref{SMN}.
 
For large $M,N$ with $M/N = \lambda$ fixed, we use
  \begin{align}
\sinh^{-1}(\sqrt \rho \sin x) &= \sqrt \rho \, x \Big[1 - \frac {1+\rho} 6 x^2
   +\frac {(1+\rho)(1+9\rho)}{120} x^4+\cdots \Big]
 \end{align}
  and \eqref{y} to obtain
 \begin{align}
  \pi y^* &= (\pi \,\tilde q)\Big[ 1- \frac {1+\rho} 6 \Big(\frac {q\pi} N \Big)^2
   +\frac {(1+\rho)(1+9\rho)}{120} \Big(\frac {q\pi} N \Big)^4+\cdots \Big]
 \end{align}
where $\bq = \lambda \sqrt \rho\, q$. 
This leads to
\begin{align}
\tanh(\pi y^*)& = \tanh (\pi \tilde q) -\frac {1+\rho} 6 (\pi \,\tilde q) \,
\sech^2 (\pi \,\tilde q) \Big(\frac {\pi q } N \Big)^2  \nonumber \\
  & +
 \Big[\frac {\pi \,\tilde q} {120} (1+\rho)(1+9\rho)  \nonumber \\
 & \quad - \frac {(\pi \,\tilde q)^2} {36} (1+\rho)^2 \tanh (\pi \,\tilde q) \Big] \sech^2 (\pi \,\tilde q)
\Big(\frac {\pi q } N \Big)^4 +\cdots. \label{tanh}
\end{align}

Substituting \eqref{tanh} into \eqref{SMN}, we obtain
\begin{align}
\Delta_{q,M,N}(\rho)
  & = D_{q,N}(\rho) \big[ \tanh (\pi \tilde q) -1 \big] \nonumber \\
 & \quad +  D_{q,N}(\rho) \times (\pi \tilde q)
  \Bigl\{   -\frac {1+\rho} 6  \,
\sech^2 (\pi \,\tilde q) \Big(\frac {\pi q } N \Big)^2 \nonumber \\
 & \quad  +(1+\rho)
{\sech^2(\pi \tilde q)} \Big[\frac{1+9\rho
}{120}-\frac{1+\rho}{36}(\pi
\tilde q)   \tanh (\pi \tilde q) \Big]  \left(\frac{\pi q}N\right)^4  \nonumber \\
& \quad + \cdots \Bigr\}.
\end{align}
Rewrite $D_{q,N}(\rho)$ given by \eqref{Deltadef} as
\begin{align}
D_{q,N}(\rho) &= \frac 2 {q\pi} + \frac 2 N f_\rho(\frac {\pi q} N) \nonumber \\
   &=\frac 1 {q\pi}\Big[ 2+ 2f_{\rho}^{(1)}(0)\Big(\frac {q\pi} N \Big)^2
  +\frac 1 3 f_{\rho}^{(3)}(0) \Big(\frac {q\pi} N \Big)^4
  + \cdots \Big]. \label{D22}
\end{align}
where the derivatives are given in \eqref{fdevs}.
This leads to the desired asymptotic expansion
  \begin{equation}
\Delta_{q,M,N}( \rho) = \sum_{i=0}^\infty \frac
{\Delta_{q,2i}(\lambda,\rho)}{N^{2i}} \label{DeltaMN}
\end{equation}
with  expansion coefficients
\begin{align}
\Delta_{q,0}(\lambda,\rho)&= \frac 2 {\pi q} \big[\tanh (\pi \bq)-1 \big], \nonumber \\
\Delta_{q,2}(\lambda,\rho)&= 2 \pi q f^{(1)}_\rho(0)  \big[\tanh (\pi \bq)-1 \big]-
\frac  {\lambda \sqrt \rho (\pi q)^2} 3
(1+\rho) \, \sech^2(\pi \tilde q), \nonumber  \\
\Delta_{q,4}(\lambda,\rho)&= \frac { (\pi q)^3 f^{(3)}_\rho(0)} 3
 \big[\tanh (\pi \bq)-1 \big] \label{DeltaResult}
 \\
 &+ \lambda \sqrt \rho (\pi q)^4(1+\rho)
\Big[\frac{53-3\rho}{180} -\frac{(1+\rho)}{18}(\pi \bq)\tanh(\pi
\bq)\Big] \, \sech^2(\pi \bq) .\nonumber
\end{align}
  As remarked earlier, values of these coefficients decrease exponentially as $q$ increases.

\section{Results}
\subsection{Summary of asymptotic expansions}
Results obtained so far may be summarised as follows: the resistance $R_{M\times N}(r,s)$
is given by \eqref{finalexpansion},  with $ R_{M\times N}(\rho)$ expanded as
 \begin{align}\label{result1}
  &R_{M\times N}(\rho)= \frac 2 \pi \log N + \sqrt \rho \Big(\frac M N
  -\frac 12\Big)+\frac 1 \pi \Big[2\gamma -1 + 2 \log \Big(\frac 2 \pi\Big)\nonumber \\
   & \quad -\log(1+\rho)  + \frac {\rho -1}
{\sqrt \rho} \tan^{-1}\sqrt \rho \, \Big]
 +L_N(\rho)  + \sum_{q=1}^{N/2} \Delta_{q, M,N}(\rho),
\end{align}
 where $\gamma=0.577\,215\,664\,901\,53\dots$ is Euler's constant,
 $L_N(\rho)$ is given by \eqref{L}  and $\Delta_{q,M,N} (\rho)$ given by \eqref{DeltaMN}.

As $N\rightarrow \infty$ with $\lambda = M/N$ fixed, \eqref{result1} can be written as
\begin{align}
R_{M\times N}(\rho) &= \frac 2 \pi \log N + C(\lambda,\rho) +
\sum_{i=1}^\infty \frac{b_{2i}(\lambda, \rho)}{N^{2i}}\label{rMN} \\
\intertext{where}
 C(\lambda,\rho) &=  \sqrt \rho \Big(\lambda
  -\frac 12\Big)+\frac 1 \pi \Big[2\gamma -1 + 2 \log\Big(\frac 2 \pi\Big)\nonumber\\
  & -\log(1+\rho)
 + \frac {\rho -1} {\sqrt \rho }\tan^{-1}\sqrt \rho \Big]+\sum_{q=1}^\infty
\Delta_{q,0}(\lambda,\rho) , \nonumber \\
  b_{2i}(\lambda, \rho)&=
 -\Big(\frac {2 B_{2i}\pi ^{2i-1} } {(2i)!} \Big) f^{(2i-1)}_\rho(0)+
 \sum_{q=1}^\infty \Delta_{q,2i}  (\lambda,\rho).
 \end{align}
 Here, the
 Bernoulli numbers are
$B_2=1/6, B_4=-1/30, B_6=1/42$ ([\ref{Hi}] equation 5.8.8).
 The function $f_\rho(x)$ is defined by
 \eqref{fdef} and its first few derivatives are given in
 \eqref{fdevs}. Equation \eqref{DeltaMN} gives an expansion of
 $\Delta_{q,M,N}(\rho)$ in inverse powers of $N^2$ correct to
  $O(1 /{N^4})$ and the coefficients decay exponentially with $q$
 so that accurate results may be obtained using only the first few
 terms of the sum. This is illustrated in Table \ref{DelTab} in the
 case $\lambda=\rho=1$.

\begin{table}
{\footnotesize
\begin{center}
\begin{tabular}{|r|l|l|l|} \hline
$q$ & $2\Delta_{q,0}$ & $2\Delta_{q,2}$& $2\Delta_{q,4}$\\
\hline 1&
-0.0047465399754997281&-0.082316647898659221&0.038515173969807909\\
2& $-4.4402067094342628\,{10}^{-6}$&-0.00067582947581056974&-0.032940604383097552\\
3&$-5.5279070728467383\,{10}^{-9}$&$-2.9215219029290850\,{10}^{-6}$&-0.00060979439982744305\\
4&$-7.7422874638854272\,{10}^{-12}$&$-9.8350012986547643\,{10}^{-9}$&$-5.3264158004237130\,{10}^{-6}$\\
5&$-1.1566622761121781\,{10}^{-14}$&$-2.8935175541424704\,{10}^{-11}$&$-3.2098297733739912\,{10}^{-8}$\\
\hline
$\Sigma_q$&-0.0047509857178701073&-0.082995408760387631&0.004959416517708477\\
\hline
\end{tabular}
\end{center}
} \caption{The  coefficients $\Delta_{q, 2i}(1,1)$ in
\eqref{square}.} \label{DelTab}
\end{table}

\subsection{The case M=N, r=s=1}
For an $N\times N$ network with $r=s=1$ we have $\lambda = \rho =1$. From \eqref{finalexpansion}
and \eqref{rMN} we obtain
\begin{align}
R_{N\times N}(1, 1) &= 2 R_{N\times N}(1)\nonumber \\
    &= \frac 4 \pi \log N + c_0 +
\frac{c_2}{N^2} +\frac{c_4}{N^4} + O(\frac 1 {N^6}),\label{Rsquare}
\end{align}
{where}
\begin{align}
 c_0 & = 2 C(1,1) + 2 \sum_{q=1}^\infty
\Delta_{q,0}(1,1) \nonumber \\
& = 1 +\frac2 \pi \left[ 2\gamma -1 +\log\Big(\frac
 2 {\pi^2} \Big)\right] + \frac 4 \pi \sum_{q=1}^\infty
 \Big(\frac {\tanh  (\pi q)-1} q  \Big) \nonumber \\
& = (0.082\ 069\ 879\ 627\ 328 \cdots)  - (0.004 \ 750\ 985\ 717\ 870\ 046\ 5 \cdots) \nonumber \\
&= 0.077\ 318\ 893\ 909\ 458 \cdots , \nonumber \\
c_2 &= -2\pi B_2 f_1^{(1)}(0) + 2 \sum_{q=1}^\infty \Delta_{q,2}(1,1) \nonumber \\
    &= 0.266\ 070\ 441\ 638\ 478\  \cdots ,\nonumber \\
 c_4 &= -\frac{\pi^3 B_4}{6}f_1^{(3)}0)+ 2
\sum_{q=1}^\infty \Delta_{q,4}(1,1) \nonumber \\
  &= -0.534\ 779\ 473\ 843\ 066\ \cdots. \label{square}
\end{align}
where we have used the data in Table \ref{DelTab}. This reproduces
numerical values of the coefficient $c_0$ determined from a
differential approximant analysis [\ref{GU}] of the first 29 values
of $R_{N\times N}(1,1)$ together with a Neville table analysis
[\ref{ETB}]. Note that the correction to the dominant contribution in $c_0$ is not negligible.
We have further extended the Neville table analysis
 of [\ref{ETB}] to the next two coefficients, and obtained
results in agreement with the theoretical values of $c_2$ and $c_4$.

Finally, the asymptotic expansion \eqref{Rsquare} is to be compared to that of the resistance between
nodes $(0,0)$ and $(N-1, N-1)$ in an infinite square lattice [\ref{cserti}],
\begin{equation}
R_{N\times N,\infty}(1, 1) = \frac 1 \pi \Big[ \log N
 + \gamma + 2 \log 2 \,\Big] + \cdots. \label{InfSquare}
\end{equation}

\section*{Acknowledgments}
FYW would like to thank David Wallace for the hospitality at the Issac Newton Institute for
Mathematical Sciences where this research was initiated. We are grateful to Wentao Lu for
help in the preparation of the manuscript.

\newpage
\section*{Appendix}
In this Appendix we evaluate $S_{q,M,N}(\rho)$ given by the summation \eqref{Sqdef} using 
  the  Euler-Maclaurin sum formula
([\ref{Hi}] equation $5.8.18$)
\begin{align}
\sum_{k=0}^{r-1}& g_{k+\frac12} = \frac 1 h\int_{x_0}^{x_r} g(x) dx\nonumber\\
&- \sum_{i=1}^m
 \frac{(1-2^{1-2i})B_{2i}h^{2i-1}}{(2i)!}\Big[g^{(2i-1)}(x_r)-g^{(2i-1)}(x_0)\Big]
 +E_m(\xi_m) \label{EulerTwo}\\
\intertext{where $g(x)$ is such that $g_i = g(x_0 + ih)$, the integer $r$ is finite, and }
 E_m(\xi_m)&=-r
 \frac{(1-2^{-1-2m})B_{2m+2}h^{2m+2}}{(2m+2)!}g^{(2m+2)}(\xi_m),\quad
 x_0<\xi_m<x_r, \nonumber 
\end{align}
where $B_{2m}$ are Bernoulli numbers.

 Using \eqref{EulerTwo} with $x_0=0$, $x_r = \pi/2$, $h=\pi/ M$, $r=M/2$,
\begin{align}
g(x) = \frac 1 {\rho \sin^2 A_q+\sin^2x}, \label{gx}
\end{align}
and noting
that the odd derivatives vanish at the endpoints, we obtain
\begin{align}
S_{q,M,N}(\rho) &=\frac1{\pi}\int_0^{\pi / 2} g(x)
{dx} +E_m(q,M, N,\xi_m) \nonumber \\
&= \frac 1{2 \sqrt \rho \sin A_q\sqrt{1+ \rho \sin^2
A_q}}+E_m(q,M, N,\xi_m).
  \label{qmn}
\end{align}
Comparison of \eqref{qmn}  with \eqref{SQ} indicates that the dominant 
contribution of $S_{q,M,N}(\rho)$   is precisely
\eqref{SQ} with $\tanh(\pi y^*)$ replaced by $1$, a result we quoted earlier.
It also identifies the error term to be
\begin{equation}
E_m(q,M,N,\xi_m)
=  \frac {\tanh (\pi y^*) -1 } {2 \sqrt \rho \sin A_q \sqrt{1+\rho\sin^2A_q} },\label{error}
\end{equation}
a result which cannot be deduced from the Euler-Maclaurin formula.
We point out that since  the denominator of (\ref{gx}) can be very small  for  $q$ and  $x$ small,
$E_m(q,M,N,\xi_m)$ does not necessarily vanish even in the limit of $m\to \infty$.

\newpage
{\Large{\bf References}}

\begin{enumerate}

\item
\label{kirch}  Kirchhoff G., 1847,
Ann. Phys. und Chemie, {\bf 72},
497-508 (1847).

\item
\label{pol} van der Pol, B., 1959,
 Lectures in Applied Mathematics,
Vol. 1, Ed. M. Kac (Interscience Publ. London) pp. 237-257.

\item
\label{ds}  Doyle, P. G. and J. L. Snell, Random walks and electric networks,
The Carus Mathematical Monograph, Series 22 (The Mathematical Association
of America, USA, 1984), pp. 83-149.

\item
\label{cserti}
Cserti, J., 2000,
 Am. J. Phys. {\bf 68}, 896-906.

\item
\label{DK}
Domany, E. and W. Kinzel, 1984 Phys. Rev. Lett. {\bf 53}, 311-4.

\item
\label{GU} Guttmann, A. J. 1989, {\it Phase Transitions and Critical
Phenomena}, Vol. 13,
    Ed. C. Domb and J. L. Lebowitz, Academic Press, 1-229.

\item
\label{ETB}
Essam, J. W., D. TanlaKishani and F. M. Bhatti, unpublished report available
at the website http://personal.rhul.ac.uk/uhah/101/.

\item
\label{wu}
Wu, F. Y., 2004, J. Phys. A: Math. Gen. {\bf 37}, 6653-73.

\item
\label{Hi} Hildebrand, F. B., 1956, {\it Introduction to Numerical
Analysis}, Tata McGraw-Hill publishing company, Bombay-Dehli.


\end{enumerate}

\end{document}